\documentclass[reprint,superscriptaddress,aps,longbibliography]{revtex4-2}
\usepackage[utf8]{inputenc}
\usepackage[T1]{fontenc}
\usepackage{amsmath}
\usepackage{amssymb}
\usepackage{graphicx}
\usepackage[capitalize, nameinlink]{cleveref}
\usepackage{subfigure}
\graphicspath{{figures/}}
\usepackage{adjustbox}
\usepackage{booktabs}
\usepackage{multirow}

\usepackage{color}
\usepackage{rotating}
\usepackage{ulem}

\begin{document}

\title{Symmetry-broken magneto-toroidal artificial spin ices: magnetization states and dynamics}

\author{Ghanem Alatteili}
\author{Lawrence A. Scafuri}
\author{Ezio Iacocca}
\affiliation{Center for Magnetism and Magnetic Nanostructures, University of Colorado Colorado Springs, Colorado Springs, CO, USA}

\date{\today}

\begin{abstract}
Magneto-toroidal artificial spin ices (MT-ASIs) are arrangements of nanomagnets that exhibit spontaneous toroidization. A ferrotoroidic order could have implications on the propagation of spin waves through this artificial spin ice, including the development of topological edge modes. Here, we numerically investigate the magnetization dynamics of an MT-ASI with and without spatial symmetry breaking. Through micromagnetic simulations, we compute the energies and ferromagnetic resonance spectra of the four lowest-order states, which exhibit ferrotoroidicity, antiferrotoroidicity, and no toroidicity. As expected, we find that the  resonant modes split when spatial symmetry is broken. To determine whether our system exhibits topologically protected edge modes, we perform semi-analytical calculations to first estimate the ferromagnetic resonance and then compute the band structure. Our results show that symmetry-broken MT-ASIs are reconfigurable by magnetic field protocols, and that their band structures depend on magnetization state.  Calculation of the Chern number indicates that the bands are topologically trivial in all cases, suggesting that the dynamic magnetic coupling is weak. The absence of a non-zero Chern number is proof of the weak dynamic coupling in ASIs, which must be addressed to unlock their full potential in magnonics applications.
\end{abstract}
\maketitle

\section{Introduction}

Artificial spin ices (ASIs) are lithographically patterned lattices composed of thin, elongated magnetic nanomagnets with well-defined easy axes~\cite{Heyderman2013,Nisoli2013,Skjaervo2020}. By positioning the nanomagnets in geometric arrays, ASIs mimic the magnetic frustration observed in spin ices. For example, the square~\cite{Wang2006} and Kagome~\cite{Qi2008} lattices represent crystal planes of a pyrochlore spin ice. In the square lattice, the ``ice-rule'' applies at each vertex~\cite{Wang2006}; while in the Kagome lattice, three nanomagnets interact at each vertex, leading to long-range frustration~\cite{Tanaka2006,Rougemaille2011,Mengotti2011,Arnalds2013}.

Because nanomagnets may be placed arbitrarily on a surface, studies of ASIs have moved beyond representations of crystal planes to investigate frustration more broadly~\cite{Nisoli2017}. Consequently, many ASI structures have been introduced, such as the pinwheel~\cite{Gliga2017,Macedo2018,Paterson2019,Li2019_pin}, trident~\cite{Farhan2017}, Shakti~\cite{Lao2018}, Santa Fe~\cite{Mondal2024}, and quadrupole~\cite{Sklenar2018} lattices. The ASI's magnetization states can be reconfigured by externally applied fields~\cite{Farhan2013}, allowing for the active manipulation of its properties. One particularly promising example is the manipulation of GHz magnetization dynamics~\cite{Gliga2013,Jungfleisch2016,Lendinez2019,Gliga2020,Lendinez2021}. For instance, by characterizing ferromagnetic resonance (FMR) responses of different states, magnetic state populations or vertex configurations may be estimated~\cite{Arroo2019,Gartside2021,Vanstone2022}. Further elaboration of this feature has enabled the use of ASIs for reservoir computing applications~\cite{Gartside2022,Saccone2022,Lee2024}.

The modification of FMR spectra also suggests changes to the dispersion relation of spin waves through the ASI. This expectation was demonstrated via semi-analytical~\cite{Iacocca2016,Iacocca2017c} and analytical~\cite{Lasnier2020,Wysin2023} models. Further progress in the field sees incursion into three-dimensional structures, including multilayers in which a soft magnetic underlayer~\cite{Iacocca2020,Negrello2022} or trilayer ASI elements~\cite{Dion2024} can dramatically increase the coupling; and three-dimensional structures~\cite{Perrin2016,May2019,May2021,Sahoo2021,Cheeninkundil2023,Saccone2023b,Guo2023,Berchialla2024}. Due to the large increase in the number of available geometries, a generalized semi-analytical model that approximately computes the spin wave band structure, G\ae{}nice, has been recently released~\cite{Alatteili2023} and applied to a three-dimensional square ice~\cite{Alatteili2024} and trilayer-based ASI experiments~\cite{Dion2024}. Despite being based on an approximate energy landscape, G\ae{}nice is able to capture the main qualitative features of the magnon band structure in ASIs.

Among the many ASI geometries, the magneto-toroidal artificial spin ice (MT-ASI) was introduced to establish ferrotoroidicity~\cite{Lehmann2019}. This geometry features proximal nanomagnets forming ``plaquettes'' located at the sides of a square lattice. Eight nanomagnets thereby interact at each vertex. Ferrotoroidicity is achieved by the circulation of magnetic moments relative to the nanomagnets' spatial positions. The authors of Ref.~\cite{Lehmann2019} indeed showed that regions of the MT-ASI exhibited a well-defined toroidization. Ferrotoroidicity was also recently observed in a Kagome dimer lattice~\cite{Li2022}, a Kagome lattice~\cite{Yue2024}, and a colloidal Cairo lattice~\cite{RodriguezGallo2023}. It is natural to inquire whether a toroidal moment may influence the GHz dynamics of ASIs, and whether such dynamics can be reconfigured by simple field protocols.
\begin{figure}[b]
\centering
\includegraphics[width=3.3in]{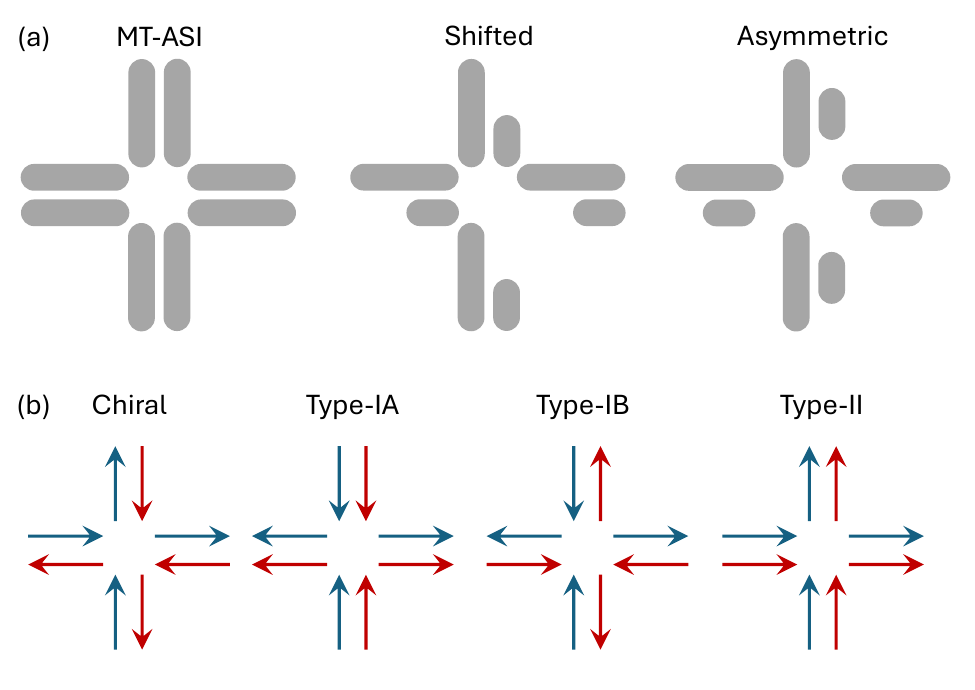}
\caption{(a) MT-ASI geometries under consideration. (b) Lowest energy states, where blue and red arrows indicate the magnetization orientation in each plaquette.}
\label{fig:sch}
\end{figure}
\begin{figure*}[t]
\centering
\includegraphics[width=6.5in]{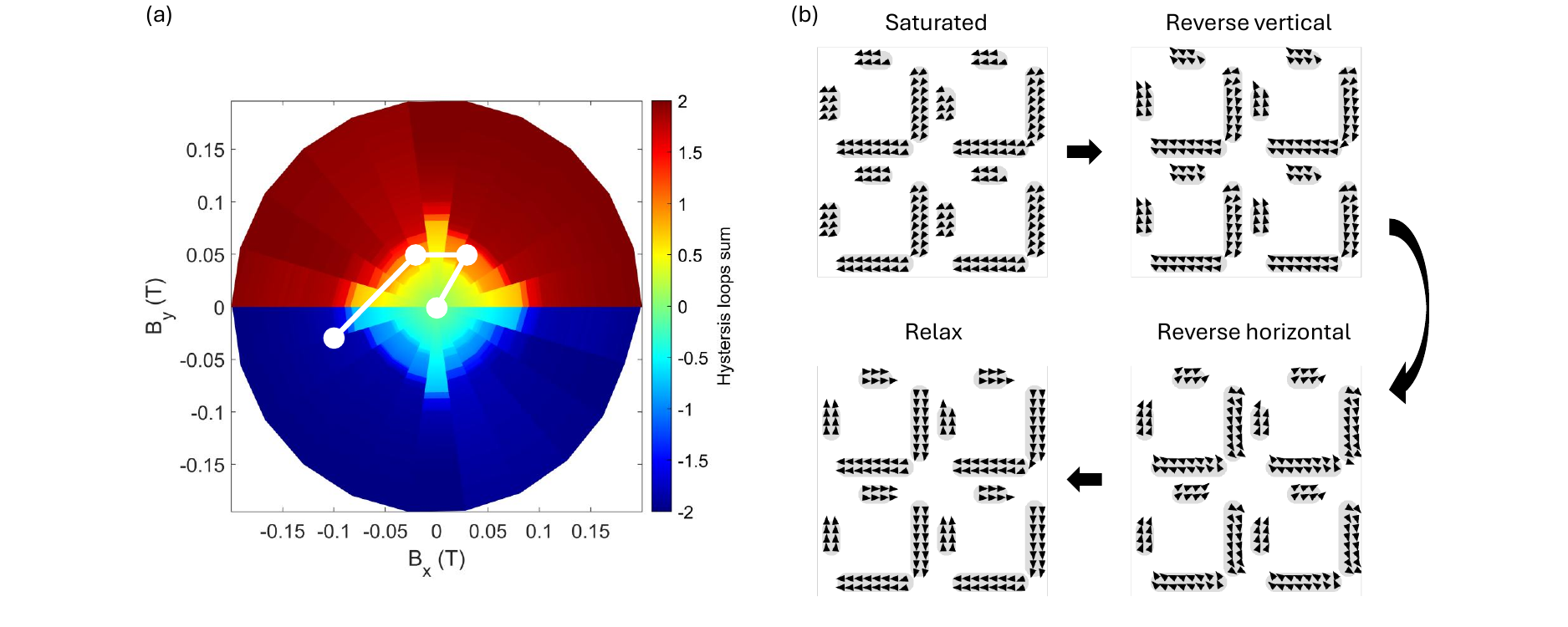}
\caption{(a) Magnetization states as a function of applied field magnitude and angle. The color scale represents the addition of the hysteresis loops for the magnetization collinear with the applied field. The white line and circles represent an applied field protocol that achieves the chiral state. (b) Progression of the magnetization states by following the field protocol illustrated in (a).}
\label{fig:poke}
\end{figure*}
\begin{figure*}[t]
\centering
\includegraphics[width=6.7in]{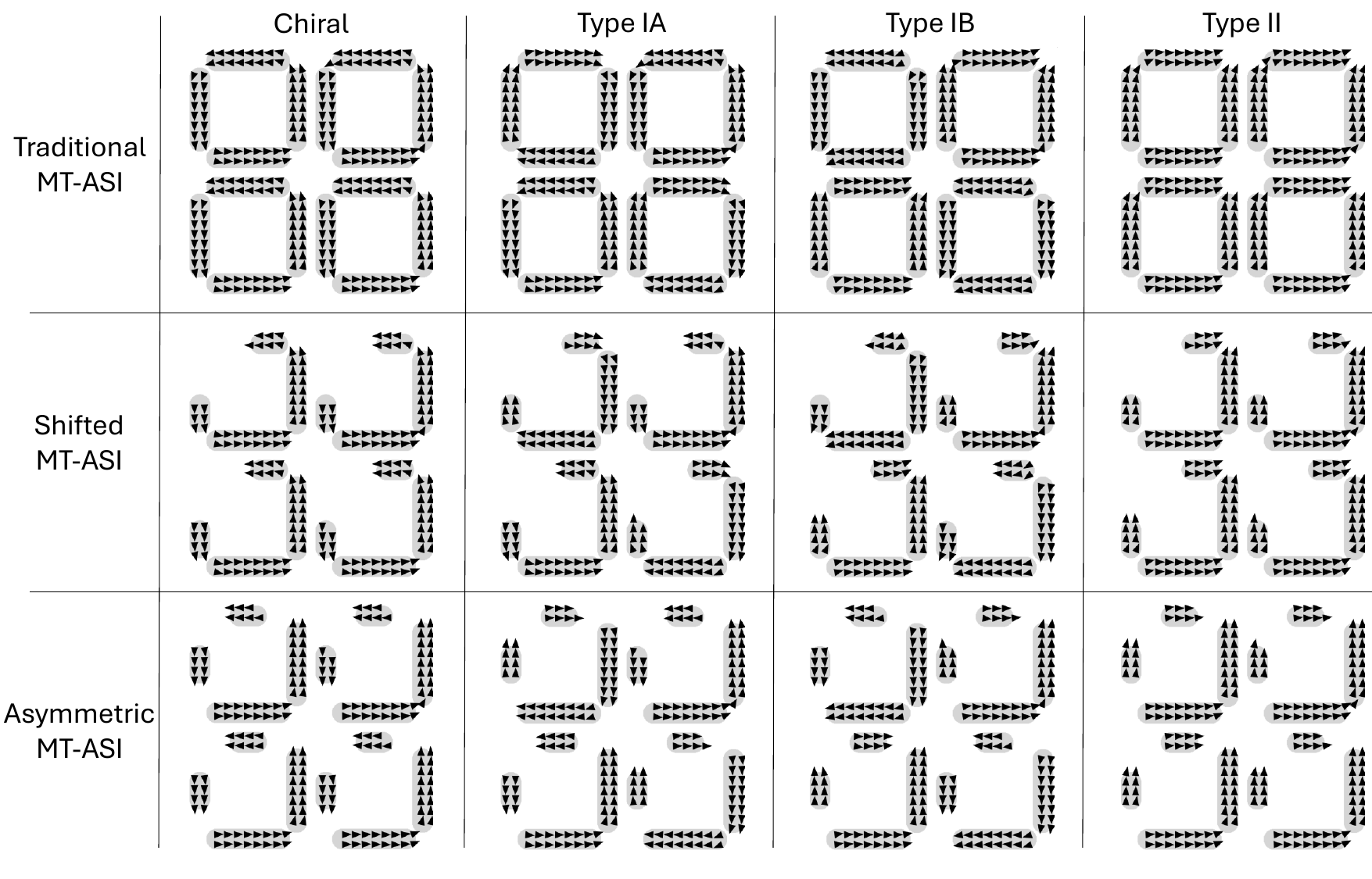}
\caption{Magnetization states in all MT-ASI configurations considered. The nanoislands are represented in gray, and the in-plane magnetization is shown by black arrows.}
\label{fig:states}
\end{figure*}

Here, we model and simulate the GHz dynamics of MT-ASIs, including lattices with spatial symmetry breaking, so that all possible magnetization states can be accessed. In section II, we investigate via micromagnetic simulations the magnetization states and the possibility of accessing them via external magnetic field protocols. The FMR is then computed by micromagnetic simulations in Section III for each magnetization state and MT-ASI structure. As expected, symmetry breaking lifts the degeneracies, thereby making more dynamical modes accessible. In Section IV, we use G\ae{}nice to compute the band structure. Comparison with micromagnetic simulations that mimic a conservative system show good agreement at the $\Gamma$ point, i.e., FMR. The band topology is then investigated via the computation of the Chern number. We find that the bands are topologically trivial in all cases. This suggests that the dynamic coupling between nanoislands is too weak in these structures to induce band topology. Finally, we provide our concluding remarks in section V.
\begin{figure*}[t]
\centering
\includegraphics[width=6.5in]{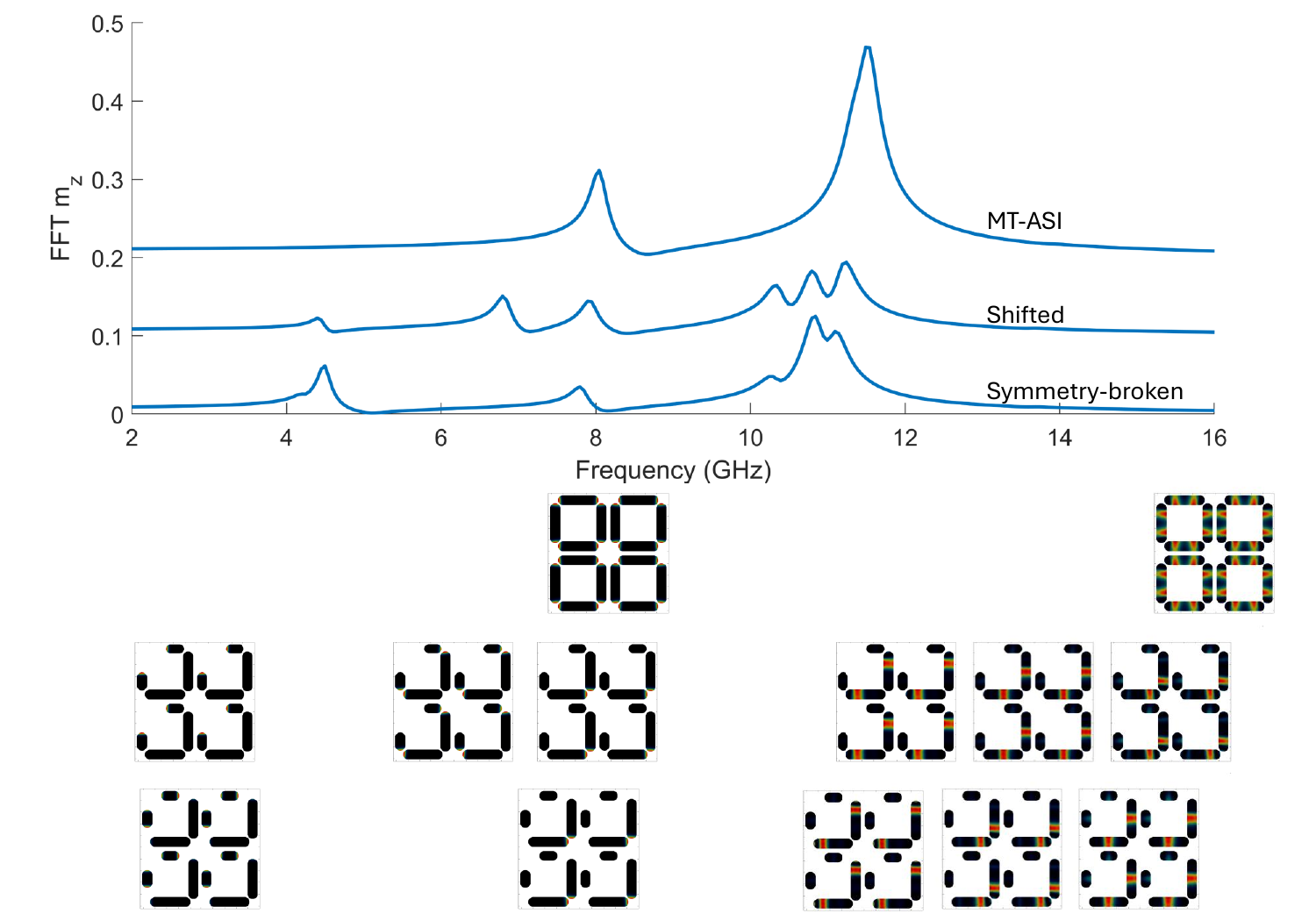}
\caption{FMR response in the chiral state for all MT-ASI geometries. The modes for each case are shown under the FMR plot and are located according to their frequency. }
\label{fig:chiralfmr}
\end{figure*}
\begin{figure*}[t]
\centering
\includegraphics[width=6.5in]{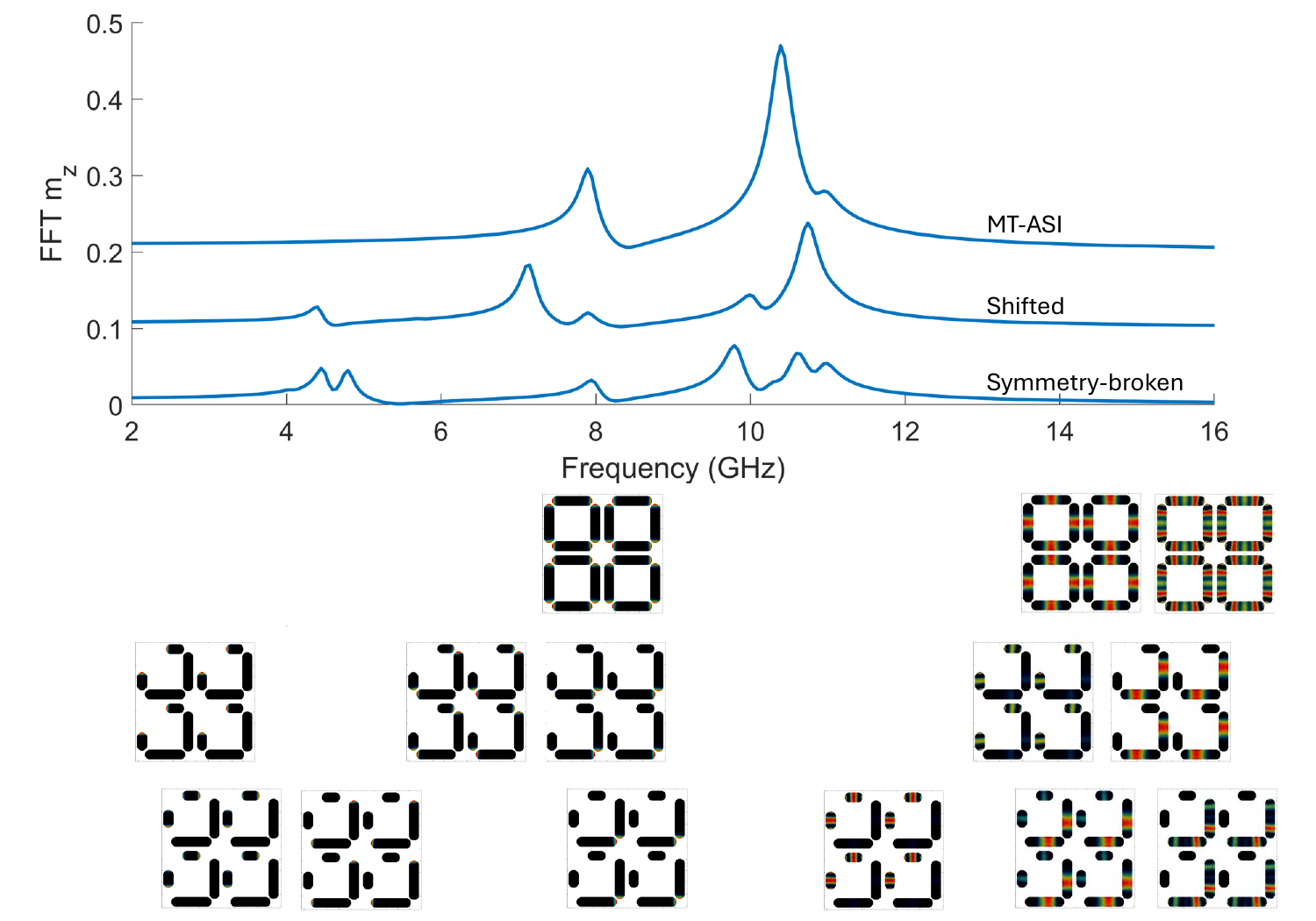}
\caption{FMR response in the Type-IA state for all MT-ASI geometries. The modes for each case are shown under the FMR plot and are located according to their frequency. }
\label{fig:typeiafmr}
\end{figure*}
\begin{figure*}[t]
\centering
\includegraphics[width=6.5in]{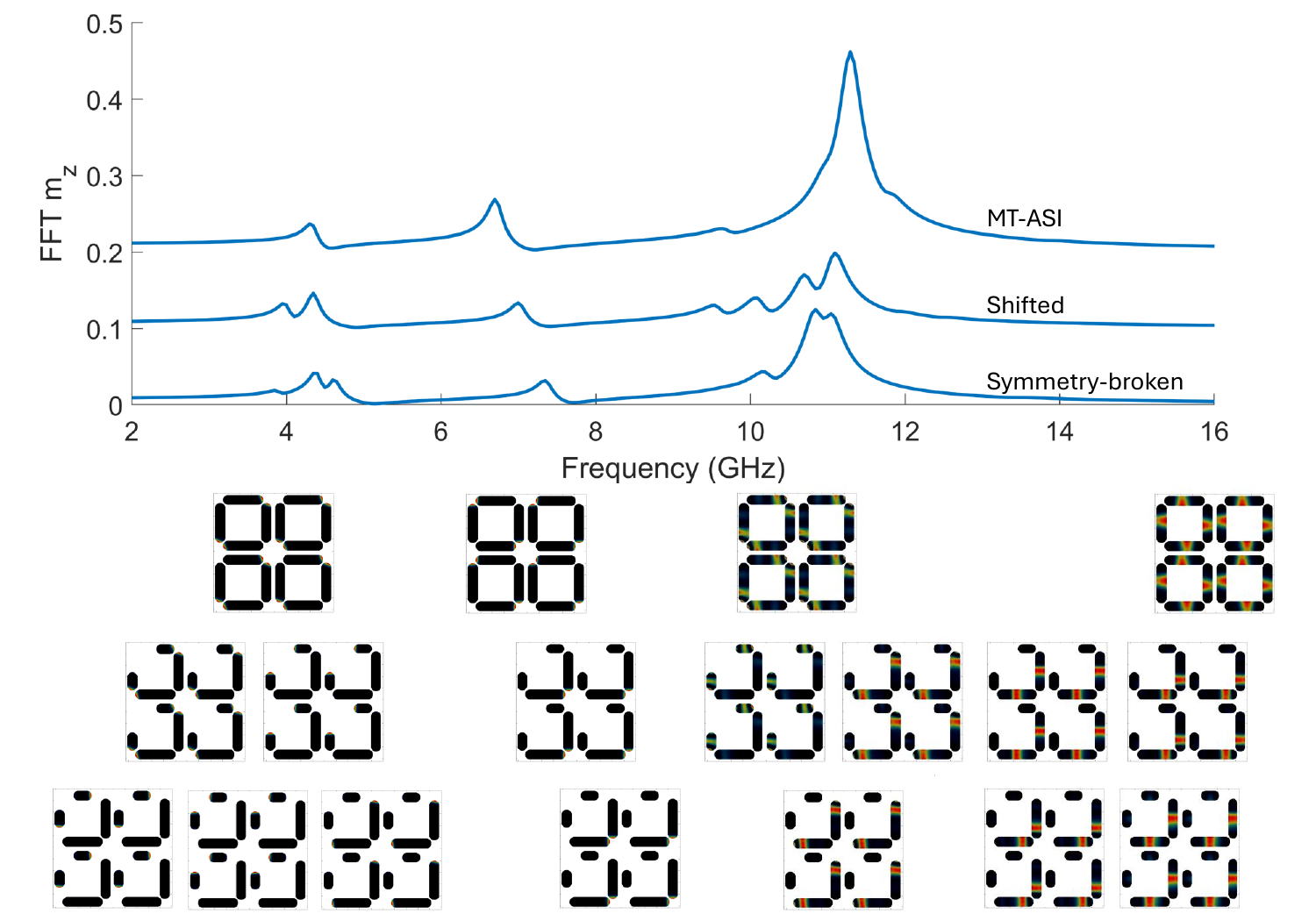}
\caption{FMR response in the Type-IB state for all MT-ASI geometries. The modes for each case are shown under the FMR plot and are located according to their frequency. }
\label{fig:typaibfmr}
\end{figure*}
\begin{figure*}[t]
\centering
\includegraphics[width=6.5in]{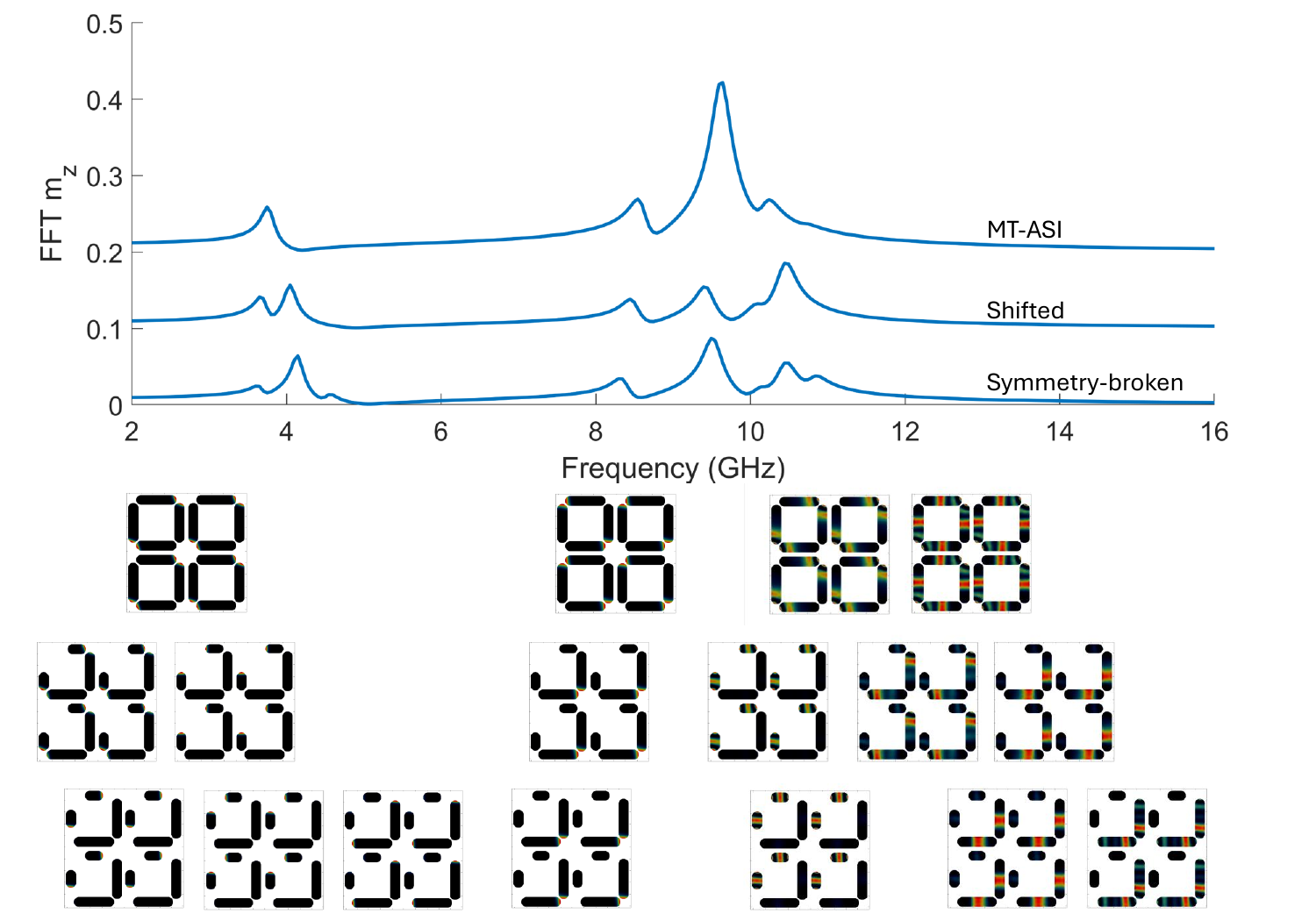}
\caption{FMR response in the Type-II state for all MT-ASI geometries. The modes for each case are shown under the FMR plot and are located according to their frequency. }
\label{fig:typeii}
\end{figure*}

\section{MT-ASI geometries and magnetization states}

We consider three geometries illustrated in Fig.~\ref{fig:sch}(a). In the leftmost, all elements are equal in size. In the two other configurations, spatial symmetry is broken by shortening the length of one nanoisland in each plaquette: in the ``shifted'' MT-ASI, the shorter elements are located close to the position of one vertex; and in the ``asymmetric'' MT-ASI, the shorter elements are positioned at the center of the plaquette. The shorter elements have a different coercivity, and their magnetization states can be accessed by external magnetic field protocols. In addition, by shortening rather than widening~\cite{Dion2019} the nanomagnets, we ensure that the center-to-center distance is kept constant.

The MT-ASI can be considered to be a square ice with a two-nanoisland plaquette, similar to the quadrupole lattice ASI~\cite{Sklenar2018}. This implies that eight states are available. We will focus on the four lowest level states akin to Type-I and Type-II states for the square ice. The four possible states are shown in Fig.~\ref{fig:sch}(b), where each sublattice is represented by a pair of blue and red arrows composing a plaquette. The ground state is the ``chiral'' state in which all the magnetic moments in the plaquettes are antiparallel and each sublattice is in a Type-II state. While seemingly counterintuitive, the chiral state minimizes total energy because the edge spins favorably close stray fields within each plaquette. The chiral state is therefore a ferrotoroidal phase. There are two varieties of Type-I states, Type-IA and Type-IB, depending on the relative orientation of the magnetization within each plaquette. Type-IA is anti-ferrotoroidal while Type-IB has no toroidization. In these cases, an antiparallel orientation of the magnetization in each plaquette has a higher energy because the stray fields are closed with interactions between different plaquettes. The Type-II state features parallel magnetizations within each plaquette, as well as a Type-II sub-lattice orientation. As for Type-IB, there is no toroidization in this state.

To access these states, the coercivities of the nanomangets must be sufficiently different. We constrained our numerical study to geometries in which the largest nanoislands have dimensions of 200~nm~\(\times\)~50~nm~\(\times\)~10~nm. The magnets within the plaquettes were separated by $s=20$~nm, and the lattice constant defined as the center-to-center distance between plaquettes was $d=300$~nm. The resulting micromagnetic unit cell~\cite{Gliga2020} was 600~nm~\(\times\)~600~nm~\(\times\)~10~nm discretized in 512~\(\times\)~512~\(\times\)~1 cells, resulting in cells of dimensions 1.17~nm~\(\times\)~1.17~nm~\(\times\)~10~nm. Despite the cells' large aspect ratio, it has previously been shown that numerical simulations using Mumax3 are sufficiently accurate to determine the resonance frequencies~\cite{Iacocca2016,Iacocca2017c,Lendinez2023,Vanga2025}. We also applied periodic boundary conditions in the $x$ and $y$ directions, setting the dipole kernel to consider 10 repetitions while free-spin boundary conditions are considered along the $z$ axis.

We reduced the length of one of the nanoislands in each plaquette until the maximum difference in coercivity was achieved. We found that nanoislands with dimensions 90~nm~\(\times\)~50~nm~\(\times\)~10~nm resulted in a coercivity difference of $\mu_0H_c=75$~mT. This coercivity is maintained throughout the ASI structure, because it depends primarily on the coupling between nanoislands within a plaquette. For example, the hysteresis loops were simulated as a function of applied field angle for the asymmetric MT-ASI. To summarize the results, we added the magnetization loops obtained by sweeping the applied field up and down in magnitude. For each angle, we computed the magnetization collinear to the applied field. The results are shown as a polar plot in Fig.~\ref{fig:poke}(a). In this representation, deep red (2) or blue (-2) colors indicate that the magnetization collinear with the applied field direction have positive or negative signs, respectively. Close to the center of the plot, transitions occur. \cite{Lehmann2019}

From this phase diagram, it is clear that specific magnetization states may be reached by reversing target sublattices. For example, we can reach the chiral (ground) state by following the path shown in Fig.~\ref{fig:poke}(a), starting from a negative saturating applied field down to zero. The magnetization states at each field transition in this protocol are shown in Fig.~\ref{fig:poke}(b). The field transitions are shown by white circles in Fig.~\ref{fig:poke}(a). To simulate this field protocol, we ramped the field in steps of 100 instances between each transition point, relaxing the magnetization state for each field change. This ensured that the time of the field sweep is not a factor in reaching the desired configuration. The protocol starts at a saturating field of $\mu_0\mathbf{H}_0=-0.1\hat{x}-0.03\hat{y}$~T. The field is then linearly changed to $\mu_0\mathbf{H}_0=-0.01\hat{x}+0.05\hat{y}$~T, leading to the reversal of the small vertical nanoislands. Then, the field is linearly changed to $\mu_0\mathbf{H}_0=0.01\hat{x}+0.05\hat{y}$~T to reverse the small horizontal nanoislands. Finally, the field is linearly changed to zero to stabilize the magnetization state. Variations of this protocol can be used to access any magnetization state as long as the spatial symmetry is broken, so that nanoislands in each sublattice have different coercivities.

Having established that the magnetization states can be accessed, we perform simulations where the states are set as initial conditions and we allow the simulation to relax in the absence of applied field. This is achieved in Mumax with the \texttt{relax()} command, in which the Larmor torque is disabled and only the dissipative term is evolved to quickly converge to the energy minimum. The relaxed states for the three MT-ASI geometries considered are shown in Fig.~\ref{fig:states}. The total energies obtained in simulation are reported in table~\ref{tab:energies}. Interestingly, the energy difference between Chiral and Type-IA states is small compared with other states.

\begin{table}[t]
\centering
\begin{tabular}{l|ccc}
       & MT-ASI & Shifted & Asymmetric \\ 
\hline
Chiral & 13.54 aJ    & 16.61 aJ   & 18.79 aJ    \\
\hline
Type IA & 13.88 aJ    & 16.85 aJ    &   19.36 aJ  \\
\hline
Type IB & 19.66 aJ    & 21.19 aJ    & 19.66 aJ    \\
\hline
Type II & 21.33 aJ   & 22.37 aJ   & 20.90 aJ   \\
\end{tabular}
\caption{Total energies for all magnetic states in the MT-ASI, shifted MT-ASI, and asymmetric MT-ASI geometries.}
\label{tab:energies}
\end{table}

\section{Ferromagnetic resonance and mode volumes}

To determine if either toroidicity or spatial symmetry breaking in MT-ASIs have any implications for their GHz dynamics, we perform simulations to obtain FMR spectra. For this, we consider the states stabilized in Fig.~\ref{fig:states} with no external field as initial states. We then apply a perturbation as a uniform field of magnitude 0.01~T in the \(z\)-direction for 10~ps. Then, we record the average magnetization evolution for the subsequent 20~ns, output in time steps of 0.025~ns. This results in a Nyquist frequency of 20~GHz and a frequency resolution of 50~MHz.

The FMR for all geometries in the chiral state is shown in Fig.~\ref{fig:chiralfmr}. The MT-ASI exhibits the typical two peaks observed in square ices, e.g.,~\cite{Lendinez2019}. As shown by the mode volumes below, the low-frequency peak at $\approx8$~GHz results from edge modes in all nanoislands, while the high-frequency peak at $\approx11$~GHz is a bulk-like mode. This high-frequency peak is rather broad, suggesting that several modes are present under its linewidth. The degeneracy of modes is broken by the other two MT-ASI geometries, shifted and asymmetric. The edge mode resonance is now split into three and two, respectively. The lowest and highest frequency edge modes in both cases are located at $\approx4$~GHz for the small nanoislands and $\approx7.9$~GHz for the large nanoislands. In addition, the shifted geometry exhibits an edge mode at the vertices where small and large nanoislands interact at $\approx6.8$~GHz. The bulk mode splits into three modes for both the shifted and asymmetric geometries. While the lowest bulk mode at $\approx10.3$~GHz exhibits a similar mode volume, the modes at $\approx10.8$~GHz and $\approx11.2$~GHz are reversed because of the effect of the small nanoislands' stray field on the large nanoislands. The split in the resonance frequencies is a clear consequence of broken symmetry.

For the Type-IA configuration, the FMR and mode volumes are shown in Fig.~\ref{fig:typeiafmr}. In this case, a clear splitting is seen for the bulk modes of the MT-ASI, with the peak at $\approx10.4$~GHz corresponding to a typical bulk mode and the peak at $\approx10.9$~GHz to a higher-order bulk mode, similar to that observed for the chiral state. For the shifted geometry, the same edge modes are observed, but there are only two prominent bulk modes for the small nanoislands at $\approx9.9$~GHz and for the large nanoislands at $\approx10.7$~GHz. For the asymmetric geometry, the lower edge modes are now split into small and large nanoislands, $\approx4.4$~GHz and $\approx4.8$~GHz respectively, and we also observe a clear peak for the small nanoislands' bulk mode at $\approx9.8$~GHz.

In the Type-IB configuration, the antiparallel orientation of the magnetization within the plaquettes results in further mode splitting. For the MT-ASI, the edge modes split into asymmetric configurations at $\approx4.3$~GHz and $\approx6.6$~GHz. This large energy split can be understood from the static configuration. In the lowest energy mode, the edge magnetization vectors repel each other so that the effective field is lower. Conversely, the higher-energy mode closes the stray field between the nanomagnets, which increases the internal field. For the shifted and asymmetric geometries, the mode splitting is similar to the Type-IA state for the edge modes and the chiral state for the bulk modes.

Finally, in the Type-II configuration, the bulk frequencies are lower than in the other states. This is understood as the Type-II state at the vertex is broken by the plaquettes so that the internal field is primarily given by the nanoislands in closest proximity. Indeed, the largest bulk mode for all cases are in the $9$~GHz range. The edge modes in all cases follow the expected behavior, i.e. mode splitting under symmetry breaking.

\section{Semi-analytical modeling}

Both the FMR and the band diagram of the MT-ASI geometries can be investigated with the semi-analytical model, G\ae{}nice~\cite{Alatteili2023}. This is a minimal model that solves for the eigenvalues of the system's Hamiltonian, so that it is computationally efficient compared with micromagnetic simulations. In prior studies, G\ae{}nice has shown good agreement with micromagnetic simulations of simple square ASI geometries~\cite{Iacocca2016,Jungfleisch2016,Iacocca2017c}. However, as the geometries of ASIs become more complex, the additional degrees of freedom result in many more frequency modes. These have often proven elusive to FMR experiments and micromagnetics simulations, possibly due to prohibited symmetries. Despite these limitations, G\ae{}nice returns reasonable qualitative predictions for complex structures, such as a trilayer-based square ASI~\cite{Dion2024} and a 3D-tilted square ASI~\cite{Alatteili2024}.

G\ae{}nice relies on the Holstein-Primakoff transformation to consider spin waves as small complex amplitudes~\cite{Slavin2009}. A minimal model is obtained by reducing each nanomagnet to a set of three macrospins coupled by exchange interactions within each nanomagnet and a dipole field between different nanomagnets. The band structure is obtained by applying Bloch's theorem in the tight-binding approximation. The resulting Hamiltonian matrix is expressed in terms of complex amplitudes~\cite{Slavin2009}, represented as an array $\underline{a}=[a_1 a_2 ... a_N]$ for $N$ macrospins in the unit cell. Therefore, the generalized Hamiltonian matrix is of dimensions $2N\times 2N$, including both $\underline{a}$ and its complex conjugate $\underline{a}^*$. The eigenvalues are obtained by solving the Hamiltonian equation
\begin{equation}
\label{eq:5}
    \frac{d}{dt}\underline{a} = -i\frac{d}{d\underline{a}^*}\begin{bmatrix}\underline{a}&\underline{a}^*\end{bmatrix}\underline{\mathcal{H}}\begin{bmatrix}\underline{a}\\\underline{a}^*\end{bmatrix},
\end{equation}
   where $\underline{\mathcal{H}}$ is divided into blocks of matrices
   \begin{equation}
 \label{eq:5_6}
     \underline{\mathcal{H}} = \begin{bmatrix}\mathcal{H}^{(1,1)}&\mathcal{H}^{(1,2)}\\\mathcal{H}^{(2,1)}&\mathcal{H}^{(2,2)}\end{bmatrix}.
 \end{equation}

Because this Hamiltonian describes spin waves, or magnons when interpreted as quasiparticles, the eigenvalues must come in pairs representing bidirectional waves. Therefore, we apply Colpa's grand dynamical matrix for bosonic excitations
  \begin{equation}
\label{eq:6}
    \omega\underline{\Psi} =\frac{\gamma}{2VM_s} \begin{bmatrix}\mathcal{H}^{(1,2)}&-(\mathcal{H}^{(1,1)})^*\\\mathcal{H}^{(1,1)}&-(\mathcal{H}^{(1,2)})^*\end{bmatrix}\underline{\Psi}=\frac{\gamma}{2VM_s}\underline{\mathcal{C}}~\underline{\Psi},
\end{equation}
 % therefore, satisfy relations relations $\mathcal{H}^{(1,1)}=(\mathcal{H}^{(2,2)})^*$ and $\mathcal{H}^{(1,2)}=(\mathcal{H}^{(2,1)})^*$ 
where $\underline{\Psi}$ is an eigenvector for each  eigenvalue $\omega$, $\gamma$ is the gyromagnetic ratio, and $V$ is the volume of each magnetic element. For simplicity, we assume that all volumes and $M_s$ are identical, but G\ae{}nice considers different macrospin volumes for the edge and bulk sections. In the model, different magnetic parameters can be set for each nanomagnet.

Because the nanomagnets are elongated, we use the Osborn approximation~\cite{Osborn1945} to determine the diagonal demag factors which is an approximation to the computation of the full demag field. A more precise calculation based on fitting micromagnetic FMR data~\cite{Martinez2023} returned little discrepancy with the demag factors. We also assume that the nanoislands' magnetization is uniform because of the small mode volume of the edge modes observed in micromagnetic simulations. Indeed, we observed negligible changes after modifying edge mode volumes and magnetization canting. All configurations have eight nanomagnets in the unit cell, and the arrays are defined by the translation vectors $\mathbf{a}_1 = d(\hat{x}+\hat{y})$ and $\mathbf{a}_2 = 2d\hat{x}$. This implies that G\ae{}nice can resolve 24 bands.

\begin{figure*}[t]
\centering
\includegraphics[width=7.5in]{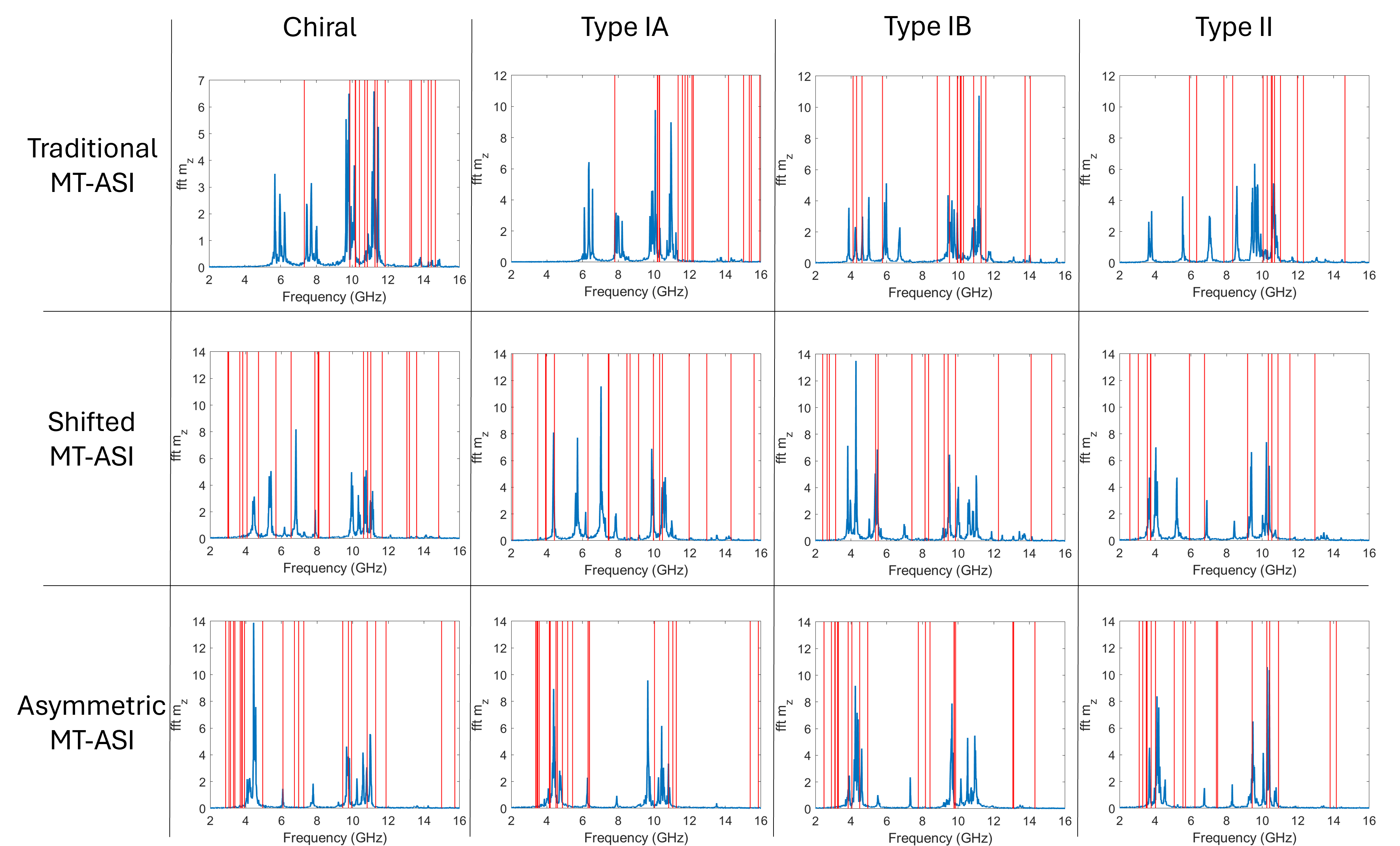}
\caption{Irreducible paths in the FBZ for traditional, shifted, and antisymmetric MT-ASI configurations for all magnetization states: Chiral, Type-IA, Type-IB, and Type-II. }
\label{fig:FMRmodesCompare}
\end{figure*}
\begin{figure*}[t]
\centering
\includegraphics[width=7.5in]{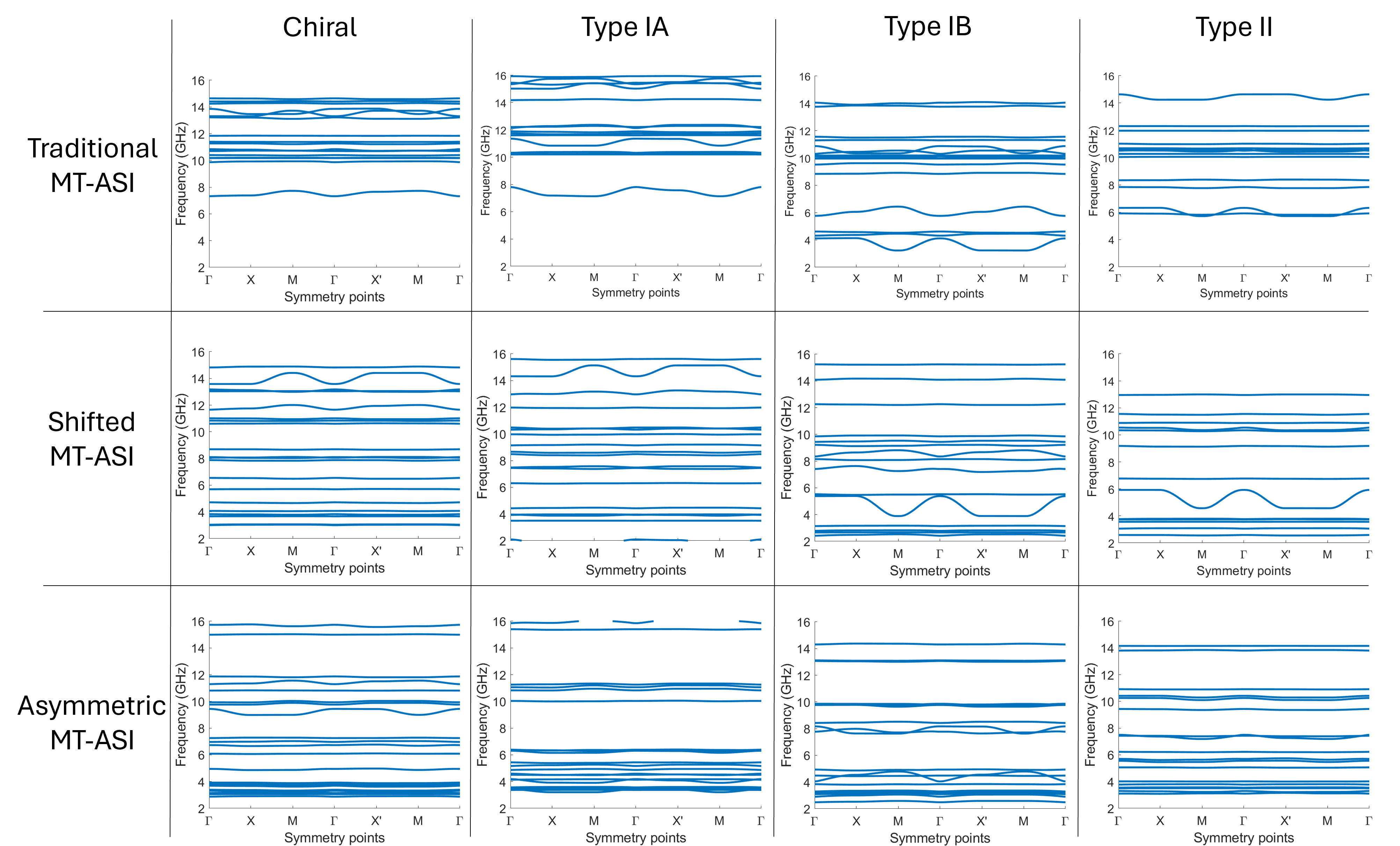}
\caption{Irreducible paths in the FBZ for traditional, shifted, and antisymmetric MT-ASI configurations for all magnetization states: Chiral, Type-IA, Type-IB, and Type-II. }
\label{fig:bandstructres}
\end{figure*}

Due to the large number of modes, we compared the results from G\ae{}nice to those from micromagnetic simulations. Clearly, from Figs.~\ref{fig:chiralfmr} to \ref{fig:typeii}, there are fewer than 24 modes. This is because of two reasons. First, we use the real damping of the assumed permalloy nanoislands, which results in a significant linewidth. Therefore, it is impossible to discern if each peak can be split into several peaks. Second, the FMR is computed as the Fourier transform of the average time signal, which corresponds to absorption experiments. To properly compare with G\ae{}nice, we need to mitigate these issues. Therefore, we perform a second set of micromagnetic simulations where we use a damping of $\alpha=0.00001$ so that the dynamics are quasi-conserved. We additionally include thermal fluctuations at room temperature (300~K) to maintain the excitation of waves with any symmetry, and we increase the simulation time to $100$~ns to improve the frequency resolution to 10~MHz. To capture all modes, we perform a Fourier transform of the magnetization time trace of each nanoisland, and then average them. This allows us to keep peaks corresponding to destructively interfering modes. The resulting FMR modes are shown in Fig.~\ref{fig:FMRmodesCompare} for all cases. We clearly see many more modes excited, which is representative of the many degrees of freedom available in micromagnetic simulations.

The $\Gamma$ point modes calculated from G\ae{}nice, corresponding to FMR, are overlaid on Fig.~\ref{fig:FMRmodesCompare} as vertical red lines. While a quantitative agreement is not expected between the two models, there are several similar features. First, the bulk-mode peaks at $\approx 10$~GHz are captured fairly well, with several modes in quantitative agreement. Second, micromagnetic simulations exhibit peaks in close proximity which are typically also captured in G\ae{}nice. Such ``mode clusters'' are due to interactions between the bulk and edge macrospins and can thus deviate considerably with edge volume and magnetization edge bending. Third, G\ae{}nice predicts several modes above $12$~GHz which are now visible in the micromagnetic simulation results, although with very low amplitude. Lastly, the peak density is well represented between the different configurations. For example, the Type-IA state in the shifted MT-ASI returns mostly equidistant modes; while in the Type-IA state in the asymmetric MT-ASI, there is a clear gap between the edge, bulk, and higher-order modes.

Having established that G\ae{}nice captures the salient features of the MT-ASI system, we now compute the full magnon band structure for all MT-ASIs cases. The band structure along the irreducible path at the first Brillouin zone (FBZ) is shown in Fig.~\ref{fig:bandstructres}. The band structures show periodic bands across all states, as required by Bloch's theorem. In general, the bands are approximately flat, with some exceptions in the Type-IB and Type-II configurations. This is a clear indication that the coupling between nanomagnets is weak, even in the traditional MT-ASI where all edges can interact. In the case of the Type-II state, the bands with stronger variation along the irreducible path originate at the vertex within plaquettes. Clearly, such variation is not available in the asymmetric MT-ASI, where the vertex coupling is the weakest. 

Despite the overall band flatness in both cases, we observe quantitative differences between the asymmetric and shifted MT-ASI cases which arise from their distinct dipole field landscape. Namely, several modes in the shifted MT-ASI are blueshifted compared with the asymmetric MT-ASI. As an example, we see in the chiral configuration of the asymmetric ASI in Fig.~\ref{fig:bandstructres} that the modes between $10$-$12$~GHz are blue-shifted by $\approx1$~GHz in the frequency spectrum of shifted MT-ASI, leading to band gaps. These are edge modes, because the modification in the band structure is only due to the different placing of the smaller nanoislands in the plaquettes. In other words, the proximity of the short nanoislands to the adjacent plaquettes in the shifted MT-ASI shown Fig.~\ref{fig:states} results in a stronger dipole interaction compared with the asymmetric MT-ASI. By contrast, modes that are negligibly affected by the change in geometry are identified as bulk modes. One example is the 11~GHz mode in the chiral state for both the shifted and asymmetric MT-ASIs.

\subsection{Band topology}

Because G\ae{}nice is optimized for computing band structures, we can investigate the existence of band topology in these MT-ASIs. Naively, the toroidicity in MT-ASI suggests potential non-trivial topological modes, particularly in the chiral state (ferrotoroidal). To identify topological phases, we compute the Chern number $C_{n}$ for band $n$, 
\begin{equation}
\label{eq:5}
    {C_{n}} =  \frac{1}{2\pi} \int_{\text{BZ}} \Omega(\mathbf{k}) \, d^2k,
\end{equation}
where \(\Omega(\mathbf{k})\) is the Berry curvature.

Because Colpa's grand dynamic matrix is Hermitian, its eigenvalues are real. Therefore, the underlying physical topology should be invariant. In other words, the Chern number derived from the Berry curvature~\cite{Fukui2005} of the spin wave band structure should be the same whether it is calculated from Colpa's eigenvalue problem or from the eigenvectors of the system’s Hamiltonian.
 
We show that the eigenvectors for a Hamiltonian $\mathcal{H}(\mathbf{k})$ and its unitary transformation \(\mathcal{H}'(\mathbf{k}) = T \mathcal{H}(\mathbf{k}) T^\dagger\) return the same eigenvalues by the unitary matrix \(T\).
Starting with the transformed Hamiltonian:
\begin{equation}
\label{eq:transf1}
\mathcal{H}'(\mathbf{k}) = T \mathcal{H}(\mathbf{k}) T^\dagger
\end{equation}

Substituting Eq.~\eqref{eq:transf1} into the eigenvalue problem for \(\mathcal{\mathcal{H}}'(\mathbf{k})\) 
%\begin{equation}
%T H(\mathbf{k}) T^\dagger |\psi(\mathbf{k})\rangle = \lambda'(\mathbf{k}) |\psi(\mathbf{k})\rangle
%\end{equation}
and multiplying both sides by \(T^\dagger\), we obtain
%\begin{equation}
%T^\dagger T H(\mathbf{k}) T^\dagger |\psi(\mathbf{k})\rangle = T^\dagger\lambda'(\mathbf{k}) |\psi(\mathbf{k})\rangle
%\end{equation}
\begin{equation}
\label{eq:eigprob}
\mathcal{H}(\mathbf{k}) T^\dagger |\psi(\mathbf{k})\rangle = \lambda'(\mathbf{k}) T^\dagger |\psi(\mathbf{k})\rangle
\end{equation}
where $\lambda'$ are the eigenvalues. Let \( |\psi(\mathbf{k})\rangle = T |u(\mathbf{k})\rangle \), then
\begin{equation}
\label{eq:eigprob2}
\mathcal{H}(\mathbf{k}) |u(\mathbf{k})\rangle = \lambda'(\mathbf{k}) |u(\mathbf{k})\rangle
\end{equation}

Equation ~\eqref{eq:eigprob2} shows that \( |u(\mathbf{k})\rangle \) is an eigenvector of \(\mathcal{H}(\mathbf{k})\) with eigenvalue \(\lambda'(\mathbf{k})\). Since \(\mathcal{H}(\mathbf{k}) |u(\mathbf{k})\rangle = \lambda(\mathbf{k}) |u(\mathbf{k})\rangle\) by definition, we conclude that
\begin{equation}
\lambda(\mathbf{k}) = \lambda'(\mathbf{k}).
\end{equation}

This demonstrates that the eigenvectors of the unitary transformed Hamiltonian \(\mathcal{H}'(\mathbf{k})\) are related to those of the original Hamiltonian \(\mathcal{H}(\mathbf{k})\) by the unitary transformation \(T\). Because Colpa's grand dynamical matrix is obtained from a unitary transformation~\cite{Colpa1978}, the computed eigenvectors $\underline{\Psi}$ from Eq.~\eqref{eq:6} can be used to recover the Chern number of the eigenvectors of the original Hamiltonian.

To numerically compute the Chern number, we use the approach by Fukui et al., ~\cite{Fukui2005}. This approach uses lattice gauge theory to calculate the Chern number in the regularly discretized FBZ, thereby reducing numerical errors that could produce non-integer results. We use G\ae{}nice to compute the band structure of all MT-ASI cases by imposing a regular grid in reciprocal space. We define a uniform square grid $21 \times 21$ with a step size of $\Delta \mathbf{k} \approx7.04\times10^{-4}$~nm$^{-1}$ along the x and y directions, totaling 441 $\mathbf{k}$-points. The Chern number is then computed using after discretizing the Berry curvature. Considering the grid points \(\mathbf{k}_{i,j}\) in the FBZ, the field strength is described in terms of the unitary link variables at each point. The link variables compute the overlap between the eigenvectors and are defined as \(U_{\mu}(\mathbf{k}_{i,j})\) along each direction \(\mu\):
    \begin{equation}
    U_{\mu}(\mathbf{k}_{i,j}) = \frac{\langle u(\mathbf{k}_{i,j}) | u(\mathbf{k}_{i+\hat{\mu},j}) \rangle}{|\langle u(\mathbf{k}_{i,j}) | u(\mathbf{k}_{i+\hat{\mu},j}) \rangle|}.
    \end{equation}

The Berry curvature is then represented as a field strength and is computed using the link variables, 
\begin{equation}
    F_{xy}(\mathbf{k}_{i,j}) = \ln \left[ U_x(\mathbf{k}_{i,j}) U_y(\mathbf{k}_{i+\hat{x},j}) U_x(\mathbf{k}_{i,j+\hat{y}})^{-1} U_y(\mathbf{k}_{i,j})^{-1} \right],
    \end{equation}
where the $-1$ superscript represents the matrix inverse. Hence, the total Chern number is obtained by summing the field strength over the Brillouin zone:
    \begin{equation}
    C = \frac{1}{2\pi i} \sum_{\mathbf{k}_{i,j}} F_{xy}(\mathbf{k}_{i,j})
    \end{equation}

Performing this computation in all 12 computed band structures returns a Chern number of zero. This implies that all bands are topologically trivial, including those in the chiral state where there is a long-range ferrotoroidicity. We believe that this is primarily due to the generally weak dynamic dipole coupling in ASIs. While stray fields strongly couple the nanoislands, resulting in the ice-rule for square ices and geometric frustration for Kagome lattices, the dynamic coupling is primarily the result of GHz excitations of edge modes. Because of their reduced mode volumes, the resulting field magnitude is weaker compared to the static stray field.

\section{Conclusions}

We investigated the magnetic states and GHz dynamics of MT-ASIs by means of micromagnetic simulations and semi-analytical calculations using G\ae{}nice. By breaking the spatial symmetry of MT-ASIs, we demonstrate that simple magnetic field protocols can be used to access all states, including the ground state which is a chiral, ferrotoroidic state. The FMR for each spatial and magnetic configuration follows expectations, in which the traditional MT-ASI exhibits degenerate bulk and edge modes while the symmetry-broken cases, shifted and asymmetric MT-ASI, break such degenerate modes.

The computations performed with G\ae{}nice returned reasonable qualitative agreement with the FMR modes. This is similar to previous investigations using G\ae{}nice~\cite{Dion2024,Alatteili2024} where the limited edge-bending discretization in the current implementation leads to quantitative disagreements. Regardless of these issues, G\ae{}nice allows one to calculate the magnon band structure of ASIs in a computationally efficient manner and much faster than what can be achieved by micromagnetic simulations. The most prominent feature in the band structure is that all bands are relatively flat, which is an indication of weak coupling in this system. The computation of a trivial Chern number for all bands further supports this hypothesis. We emphasize that the static coupling in ASIs is strong as it leads to different magnetization states with different energies. However, the dynamic coupling is mediated by stray fields from edge modes, which are much weaker than the static fields.

Our results suggest the need for increasing the dynamic coupling in ASIs should these metamaterials be used for magnonics or any application based on wave propagation. However, the modifications of the FMR due to magnetization states can still be utilized for FMR-based applications. In particular, our proposed symmetry-broken states allows for reconfigurable magnetization states insofar as a 2D magnetic field is available, as with a projected field electromagnet, for example. While our simulations suggest magnetic fields in the $150$~mT range, larger nanoislands will require stronger fields. This might be accompanied by a larger variation in the coercivity between the large and small nanomagnets in each plaquette, which is also beneficial when considering roughness in the lithography process.

\section*{Acknowledgments}

The authors thank Prof. Justin Cole for fruitful discussions regarding the calculation of Chern numbers. This material is based upon work supported by the National Science Foundation under Grant No. 2205796.

\section*{Data Availability Statement}

The data that support the findings of this article are openly available~\cite{OSF_NSF}.

\end{document}